\documentclass[conference]{IEEEtran}
\IEEEoverridecommandlockouts
\usepackage{cite}
\usepackage{amsmath,amssymb,amsfonts}
\usepackage{algorithmic}
\usepackage{graphicx}
\usepackage{textcomp}
\usepackage{xcolor}
\usepackage{subcaption}
\usepackage{booktabs}
\usepackage{url}
\def\BibTeX{{\rm B\kern-.05em{\sc i\kern-.025em b}\kern-.08em
    T\kern-.1667em\lower.7ex\hbox{E}\kern-.125emX}}

\usepackage{etoolbox} % Required for changing the bibliography font size

\AtBeginEnvironment{thebibliography}{\small} % Change the font size of the bibliography

\begin{document}

\title{On mission Twitter Profiles: A Study of Selective Toxic Behavior}
\author{
  \IEEEauthorblockN{
    \begin{tabular}{c}
      1\textsuperscript{st} Hina Qayyum \\
      \textit{Macquarie University} \\
      Sydney, Australia \\
      hina.qayyum@students.mq.edu.au
    \end{tabular}
    \hspace{2em}
    \begin{tabular}{c}
      2\textsuperscript{nd} Muhammad Ikram \\
      \textit{Macquarie University} \\
      Sydney, Australia \\
      muhammad.ikram@mq.edu.au
    \end{tabular}
    \hspace{2em}
    \begin{tabular}{c}
      3\textsuperscript{rd} Benjamin Zi Hao Zhao \\
      \textit{Macquarie University} \\
      Sydney, Australia \\
      ben\_zi.zhao@mq.edu.au
    \end{tabular}
  }

  \vspace{1.5em}

  \IEEEauthorblockN{
    \begin{tabular}{c}
      4\textsuperscript{th} Ian D. Wood \\
      \textit{Macquarie University} \\
      Sydney, Australia \\
      ian.wood@mq.edu.au
    \end{tabular}
    \hspace{2em}
    \begin{tabular}{c}
      5\textsuperscript{th} Nicolas Kourtellis \\
      \textit{Telefonica Research} \\
      Barcelona, Spain \\
      nicolas.kourtellis@telefonica.com
    \end{tabular}
    \hspace{2em}
    \begin{tabular}{c}
      6\textsuperscript{th} Mohamed Ali Kaafar \\
      \textit{Macquarie University} \\
      Sydney, Australia \\
      dali.kaafar@mq.edu.au
    \end{tabular}
  }
}

\maketitle

\begin{abstract}
The argument for persistent social media influence campaigns, often funded by malicious entities, is gaining traction. These entities utilize instrumented profiles to disseminate divisive content and disinformation, shaping public perception. Despite ample evidence of these instrumented profiles, few identification methods exist to locate them in the wild.
To evade detection and appear genuine, small clusters of instrumented profiles engage in unrelated discussions, diverting attention from their true goals~\cite{phillips2013house}. This strategic thematic diversity conceals their selective polarity towards certain topics and fosters public trust~\cite{walter2020russian}.

This study aims to characterize profiles potentially used for influence operations, termed ``on-mission profiles," relying solely on thematic content diversity within unlabeled data. Distinguishing this work is its focus on content volume and toxicity towards specific themes. Longitudinal data from 138K Twitter (rebranded as $\mathbb X $) profiles and 293M tweets enables profiling based on theme diversity. High thematic diversity groups predominantly produce toxic content concerning specific themes, like politics, health, and news—classifying them as ``on-mission" profiles.

Using the identified ``on-mission" profiles, we design a classifier for unseen, unlabeled data. Employing a linear SVM model, we train and test it on an 80/20\% split of the most diverse profiles. The classifier achieves a flawless 100\% accuracy, facilitating the discovery of previously unknown ``on-mission" profiles in the wild.
\end{abstract}


\maketitle
\section{Introduction}
\label{sec:intro}
The proliferation of persistent social media influence campaigns, often funded by malicious entities, is attracting increasing attention. These entities employ instrumented profiles to disseminate polarizing content, misinformation~\cite{xia2019disinformation}, and engage in orchestrated campaigns to propagate divisive discussions~\cite{263806}. Identifying such profiles has become a research focus. \cite{doi:10.2105/AJPH.2018.304567} examines social media influence operation methods based on labeled datasets of Russian Trolls and spam bots, revealing their focus on specific topics. \cite{walter2020russian} delves into thematic personas adopted by Russian Internet Research Agency (IRA) profiles, highlighting profiles discussing politics and everyday themes to appear genuine. For instance, they explore Twitter ($\mathbb X$) \footnote{This dataset was collected during Twitter's operation, and consequently, we refer to it as ``Twitter" throughout this text.}accounts targeting potential Trump supporters by discussing themes stereotypically linked to this group. These thematic personas strategically engage unsuspecting users during elections and political campaigns. \cite{doi:10.1126/sciadv.abb5824} emphasizes influence operations involving human-like paid trolls or automated profiles. \cite{saeed2022trollmagnifier} employs machine learning to identify instrumented Twitter profiles by training a classifier on labeled datasets of Russian-sponsored troll accounts. Meanwhile, \cite{Addawood_Badawy_Lerman_Ferrara_2019} underscores the need for content-based identification of instrumented profiles based on real-world timeline data.

Although these detection techniques contribute insights, they often rely on labeled datasets and focus on state-sponsored trolls. Consequently, their performance suffers when faced with smaller, emerging troll networks, exacerbated by \cite{phillips2013house}'s report that profiles with a toxic agenda often operate in small, coordinated groups. Thus, a more generalized methodology that identifies such profiles based solely on their individual content, without relying on prior knowledge of known troll networks or campaigns, is essential.
This study addresses this gap by identifying instrumented profiles in the wild without labeled data. We propose the existence of ``on-mission profiles," which camouflage themselves as genuine through diverse thematic content, solely designed to spread a toxic narrative about certain themes such as politics and health. We isolate these profiles based on the disproportionate volume and toxicity of their tweets concerning different themes.
Next, leveraging the identified ``on-mission" profiles, we develop and apply a classifier to identify such profiles. Our approach employs a linear SVM model, trained and tested on the most thematically diverse profiles in an 80/20 split. Remarkably, the model achieves perfect accuracy (100\%) during testing.
The key contributions of this paper are as follows:
\begin{itemize}
\item We focus on ``on-mission profiles" known for consistently posting toxic or polarizing tweets on specific themes such as politics, current affairs, and health.
Utilizing longitudinal timeline data, we assess content toxicity across different themes, providing insights into the behavior of these profiles. Our analysis reveals a bursty posting pattern with intermittent periods of inactivity among these on-mission profiles.
\item Using state-of-the-art Contextualized topic modeling with BERTweet embeddings on 230,383 unique tweets, we achieve an NMPI coherence score of 0.69, capturing meaningful topics. Diverse profiles tend to post relatively more toxic tweets on specific themes.
\item We introduce a novel approach to identify ``on-mission profiles'' by assessing their thematic contribution entropy and cross-referencing them with Botometer scores for human-like validation. This yields 96 on-mission profiles, with 62 primarily involved in toxic political discourse, 26 in health, and 8 in current affairs.
\item Using content, activity, and profile-based features of identified on-mission profiles, we develop a machine learning classifier achieving F1-scores ranging from 68.5\% to 100\% and predictive accuracy from 67.7\% to 100\% in detecting on-mission profiles. Ablation studies demonstrate the complementary nature of feature categories in profile detection. The model is subsequently applied to unlabeled profiles in our dataset.
\item We validate our classifiers on a larger dataset of 98,814 profiles, manually validating them against 512 randomly selected profiles. Our classifier correctly identifies 72\% of profiles as on-mission, revealing previously unidentified profiles. We share our dataset, including tweet IDs and toxicity scores, as well as the code, to support further research in this field.
\end{itemize}

The paper is structured as follows. Section~\ref{sec:Methodology} outlines our methodology, encompassing dataset collection (cf. \S\ref{sec:Data}), tweet-level topic modeling (cf. \S\ref{sec:Topic Modeling}), categorization of topics (cf. \S\ref{sec:Topic category}), toxicity assessment (cf. \S\ref{sec:Topic toxicity}), and thematic diversity categorization (cf. \S\ref{sec:Thematic diversity}). Section~\ref{sec:Thematically diverse profile groups trends} delves into characterizing profile groups with varying thematic diversity. Section~\ref{sec:Case Study: Thematically most diverse group (Group-VIII)} investigates Group-VIII to identify on-mission profiles, leading to the proposed detection model in Section~\ref{sec:Proposed detection model}. Ethical considerations and conclusions are discussed last.
\section{Related Work}
\label{sec:related_work}
In contemporary society, social media platforms hold a significant presence, but alongside their widespread use, entities in the form of "fake," "un-authentic," or "troll" accounts~\cite{chou2009social, FORNACCIARI2018258,roy2020fake} have generated and disseminated detrimental content such as fake news~\cite{pierri2019false}, toxicity~\cite{almerekhi2019detecting,sheth2022defining}, spam ~\cite{jin2011socialspamguard,jain2019spam}, and marketing~\cite{tiago2014digital}. The surge in influence operations on social media, especially for political purposes, has led to a growing need for identifying inauthentic profiles. However, existing methods are limited in scope and specificity.

Prior studies have explored influence operations with machine learning and cognitive behavioral modeling~\cite{Ferrara_2017}, but these are constrained by small datasets. Specialized studies focus on particular operations~\cite{10.1145/3359229}, lacking a standardized methodology. Language-based identification~\cite{ghanem2019textrolls} has also been attempted. These approaches do not generalize well and struggle with small, coordinated troll groups~\cite{farkas2020disguised}. Disguised propaganda often stems from small partisan groups, making detection challenging~\cite{phillips2013house}.
To address these gaps, this study identifies ``on-mission profiles" that consistently post toxic content on specific themes, often evading detection. We extend existing content-based methods~\cite{10.1145/3308560.3316495} by considering content diversity and toxicity across themes. Network-based approaches~\cite{makkar2019cognitive} miss solitary on-mission profiles.

Other works target specific behaviors like spam~\cite{roy2020deep,chen2015trusms,ercsahin2017twitter} or online bullying~\cite{konikoff2021gatekeepers,jha2020dhot,lee2018abusive}, but overlook profiles that disproportionately post toxic content across multiple topics. Closest to our study,~\cite{10.1093/pubmed/fdaa049} qualitatively analyzes world leaders' tweets, but lacks a generalized methodology and longitudinal perspective.
Our approach aims to bridge these gaps by focusing on identifying on-mission profiles without labeled data. We hypothesize that toxic posts on specific topics signal compromised profiles. By extending this hypothesis, we build a classifier that differentiates on-mission profiles. Our methodology leverages longitudinal data and content diversity, enhancing the detection of such profiles across various themes.

In conclusion, this work addresses the limitations of current methodologies for identifying inauthentic profiles on social media platforms. By focusing on content diversity and toxicity, we contribute a more generalized approach for uncovering on-mission profiles that consistently post toxic content across multiple themes, filling a significant gap in the field.

\section{Methodology}
\label{sec:Methodology}
\subsection{Data}
\label{sec:Data}
We make use of seven publicly available datasets that investigate misbehavior on Twitter, including \cite{gomez2019exploring,kaggle:metoomovement,ribeiro2018like,founta2018large,jha-mamidi-2017-compliment,waseem-hovy-2016-hateful}, and \cite{waseem-2016-racist}. In total, we gather 143K English profile IDs from these initial datasets.
Subsequently, we retrieve longitudinal timeline data for each profile ID using Twitter's API, capturing the 3,200 most recent tweets (owing to limitations in the Twitter API). After excluding profiles with fewer than 10 tweets, we retain 138,430 Twitter profiles. This accumulation spans approximately 293 Million tweets over 16 years, ranging from 2007 to 2021. On average, each timeline consists of 2,051 total tweets and 1,160 unique tweets (Numerous tweets are verbatim repetitions).
In addition to retrieving the actual tweet content, we also gather auxiliary tweet components, such as hashtags, URLs, and profile metadata. This metadata includes names, descriptions, mentions, and other relevant information.
%-----%
\subsection{Topic Modeling}
\label{sec:Topic Modeling}
The primary goal of this paper is to detect on-mission profiles based on the relative toxicity and volume of their tweets across different themes. For this, we employ topic modeling to identify topic categories or themes within a profile's timeline. This involves utilizing a contextualized topic model (CTM) \cite{bianchi-etal-2021-pre}, where individual tweets serve as documents and 200 topics are predefined for modeling. \emph{CTM} is a recent neural topic model architecture known for its top-notch performance on topic coherence metrics. To input the CTM model, we use BERTweet \cite{bertweet} embeddings with a vocabulary size of 5000. \emph{BERTweet} is a language model pre-trained on 850M English tweets, excelling in tweet-based NLP tasks.

We start by utilizing a general-purpose pre-trained model for our topic model. Subsequently, we fine-tune this model using the OCTIS algorithm~\cite{terragni-etal-2021-octis} with our tweet data to determine optimal parameters. To achieve this, a random 10\% sample from our corpus is employed, yielding 230,383 unique training tweets after undergoing BERTweet preprocessing.
We eliminate retweets and duplicate tweets, then implement the BERTweet pre-processing, which includes substituting user mentions and URLs with special tokens (``@USER'' and ``HTTPURL''). Emojis are converted to text strings, and tweets with fewer than 10 or more than 64-word tokens are excluded. After fine-tuning, our model achieves a topic NPMI coherence of 0.69 on a separate hold-out dataset. Subsequently, we apply this trained model to the complete tweet corpus.

\subsection{Topic category}
\label{sec:Topic category}
The topics extracted by the CTM model often lack direct interpretability and can exhibit higher-level correlations. To enhance clarity, we manually label the 200 topics into categories before assessing their toxicity.

Our proposed process for categorizing topics begins by examining the 30 most prominent words in each topic using LDAvis~\cite{pylDAvis}. This approach identifies eight main topic categories, namely ``politics," ``news/blogs," ``health/covid," ``sports," ``profanity," ``entertainment," ``everyday," and ``no topic." Three authors then assign one of these categories to each of the 200 topics. Disagreements were resolved through majority voting (2 out of 3 annotators). Fleiss's Kappa of 0.67 indicates good agreement between annotators. Here are the category specifics, ordered by topic counts and proportions:
\begin{itemize}
\item {\bf Everyday} [49 or 24.5\%]: Covers greetings, birthday wishes, pleasantries, and day-to-day discussions.
\item \textbf{No topic} [43 or 21.5\%]: Encompasses collections of unintelligible words, like prepositions and conjunctions.
\item \textbf{News/Blogs} [39 or 19.5\%]: Includes news, references to newspapers, can cover diverse unrelated topics.
\item \textbf{Politics}  [22 or 11.0\%]: Relates to politicians, government policies, underpayment, and jobs.
\item \textbf{Entertainment} [17 or 8.5\%]: Focuses on horoscopes, music, music concerts, and TV shows.
\item \textbf{Sports} [16 or 8.0\%]: Involves sports events and players.
\item \textbf{Profanity} [8 or 4.0\%]: Foul language and cursing.
\item \textbf{Health/Covid} [6 or 3.0\%]: Addresses health-related discussions, including those about Covid.
\end{itemize}
This categorization process offers a clearer overview of the diverse topics and their distribution within the dataset.
% %-----%
% \subsection{Topic toxicity}
% \label{sec:Topic toxicity}
% Now that we have applied topic modeling on the tweet level, every tweet will have a topic probability vector, with its likelihood of belonging to each topic. From this vector, each tweet can be assigned to one of the 200 topics by considering its highest probability. For example, a tweet's highest probability is in Topic 5, which is also categorized as a "politics" topic, this tweet can be generally regarded as a tweet about politics. 
% With these topics and topic categories (Section~\ref{sec:Topic category}) in mind, we augment each tweet with \emph{Perspective API}'s model of \emph{"Toxicity"}~\cite{perspective} on all the tweets. The Perspective API uses machine learning to assign a probability score ranging from 0 to 1 that measures the toxicity within a given tweet's text. Higher scores are more toxic. Overall, we query Perspective API to determine the toxicity score for 293 million tweets in our dataset. 

% To compare how toxicity is concentrated between our 200 topics, we present the median toxicity score per topic. In Figure~\ref{fig:measurement_plots/median_tox_200_topics.pdf} we can see the median toxicity per topic.
% %
% \begin{figure}[h!] \centering
% \includegraphics[width=0.90\columnwidth]{measurement_plots/median_tox_200_topics.pdf}
% \caption{A boxplot of median toxicity of 200 identified topics in topic modeling results (Section~\ref{sec:Topic toxicity}).}
% \label{fig:measurement_plots/median_tox_200_topics.pdf}
% \end{figure}
% %
%-----%
\subsection{Thematic diverse groups}
\label{sec:Thematic diversity}
The establishment of our topics and topic categories leads us to organize our profiles into groups based on the number of different categories they engage with and post about. Profiles with a higher number of categories exhibit greater thematic diversity. This diversity, coupled with the toxicity levels in their Twitter timelines, enables us to assess if they display partial or toxic behavior in specific themes.
Regarding the identified categories, we calculate the Category Probability (CP) for each profile within each category using the formula:

\begin{equation}  
CP = \frac{\texttt{Tweets per topic category}}{\texttt{Total profile tweets}}.
\label{eq:tp}
\end{equation}

A profile's \emph{Category Probability Vector (CPV)} represents the probabilities of posting in each of the 8 categories:

\begin{equation}
CPV = [CP_{(C_1)},~CP_{(C_2)}~...CP_{(C_8)}].
\end{equation}

Subsequently, we compute the Shannon Entropy~\cite{61115} for the CPV of all 138K profiles. The \emph{Shannon Entropy} quantifies the distribution of posts across different categories, calculated as $H = -\sum_C P(C) * \ln(P(C))$. Lower entropy indicates skewed posting behavior across categories, while higher entropy indicates balanced distribution.

\begin{figure}[t!] \centering
\includegraphics[width=0.95\columnwidth]{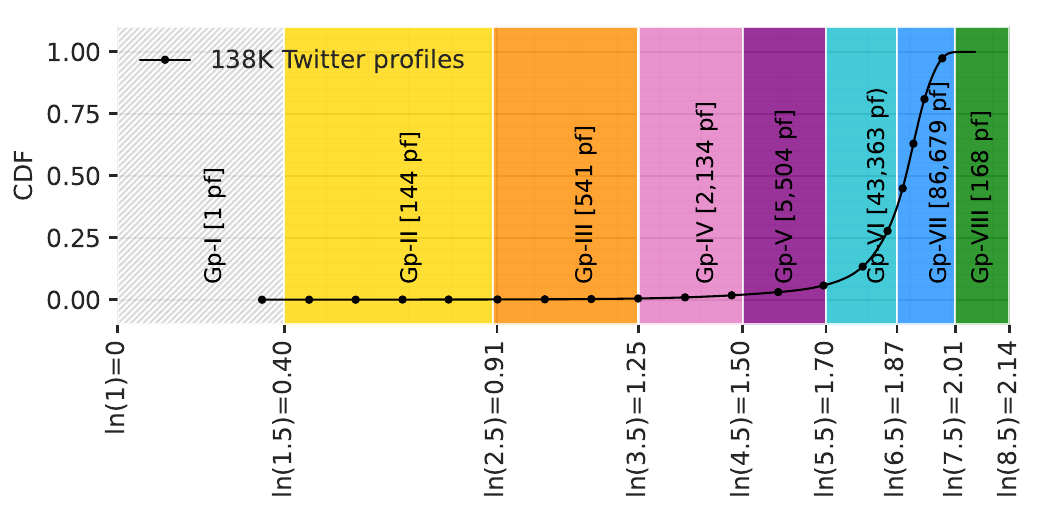}
\caption{\centering \small Entropy (H) of Category Probability Vector (CPV) per Twitter profile, organized into groups based on the number of categories identified in the tweets of these profiles (Section~\ref{sec:Thematic diversity}). Each interval represents a distinct group of profiles.}
\label{fig:entropy_bins_cdf.pdf}
\vspace{-4mm}
\end{figure}

Figure~\ref{fig:entropy_bins_cdf.pdf} depicts the cumulative distribution of entropy values per profile. We segment the entropy values into 8 intervals, selecting boundaries based on the fact that a prototype profile with uniform probability across 2 topics has an entropy of ln(2) = 0.69 and zero probability on other topics. We set boundaries at values of 1.5, 2.5, 3.5, etc., representing midpoints between whole numbers of topics. For instance, profiles in group-II discuss approximately 2 topics, resulting in an entropy range of ln(1.5)=0.40 to ln(2.5)=0.91. We continue this pattern for group-III (entropy range 0.91 to 1.25), and so forth. Thus, the number of distinct categories discussed in each profile group determines their respective names, such as group-II (2 categories), group-III (3 categories), and so on. Note that due to only one profile in group-I, it is excluded from further investigation due to limited participation in results.

\section{Thematically diverse profile groups trends}
\label{sec:Thematically diverse profile groups trends}
In this section, we analyze the 8 groups identified in Section~\ref{sec:Thematic diversity}, each comprising profiles discussing 2 to 8 distinct categories, such as politics, sports, health, etc., within their tweets. We assess the content, activity, and profile-based features of these groups in the subsequent subsections.

%-----%
\subsection{Content analysis}
\label{sec:Content based analysis}
\subsubsection{Toxicity}
\label{sec:Toxicity}
\begin{figure}[!ht]
    \begin{subfigure}[b]{.49\linewidth}
    \centering
    \includegraphics[width=\linewidth]{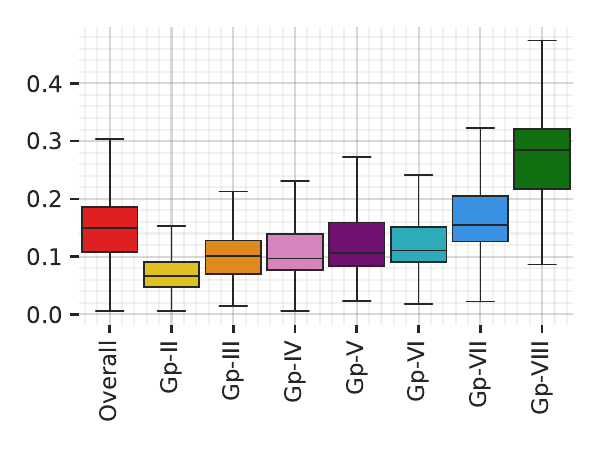}
    \caption{Toxicity scores}\label{fig:measurement_plots/TOXICITY_median_score_box_groups.pdf}
    \end{subfigure}
    \begin{subfigure}[b]{.49\linewidth}
    \centering
    \includegraphics[width=\linewidth]{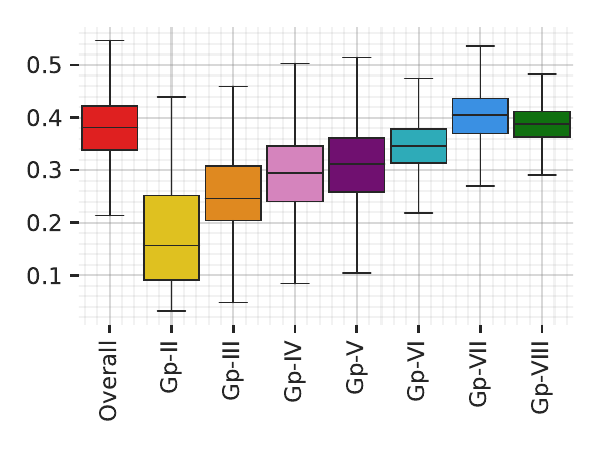}
    \caption{Gini scores}\label{fig:measurement_plots/TOXICITY_gini_score_box_groups.pdf}
    \end{subfigure}
    \vspace{-2mm}
    \caption{\centering \small Boxplots illustrating Perspective toxicity scores (\subref{fig:measurement_plots/TOXICITY_median_score_box_groups.pdf}) and the consistency of these toxic scores (\subref{fig:measurement_plots/TOXICITY_gini_score_box_groups.pdf}) for profiles within groups II-VIII (Section~\ref{sec:Topic toxicity}).}
    \label{fig:toxicity}
    \vspace{-5mm}
\end{figure}
We start by exploring the potential correlation between tweet toxicity and the diversity of thematic content within a profile. To accomplish this, we employ Google's Perspective score for "Toxicity" which analyzes the text of each profile's tweets~\cite{perspective}, as discussed in Section~\ref{sec:Topic toxicity}. By utilizing toxicity scores for each tweet, we gauge the concentration of toxicity within a profile's tweets using the Gini Index.
The Gini Index, typically employed to measure wealth inequality, captures the concentration of a certain attribute within a dataset. A Gini value closer to 0 signifies a more consistent attribute distribution, while a value approaching 1 indicates a wider range. Applying this concept to tweet toxicity scores, we compute and compare the Gini values across profiles.
The distribution of both the median toxicity scores and Gini Index values is illustrated in Figure~\ref{fig:measurement_plots/TOXICITY_median_score_box_groups.pdf} and \ref{fig:measurement_plots/TOXICITY_gini_score_box_groups.pdf}. The leftmost red boxplot in both figures serves as a benchmark, aggregating profiles across all groups.
From the figures, it's evident that thematic diversity relates positively to content toxicity. Groups VI-VIII exhibit the highest median toxicity scores (0.18 to 2.9), while group-II has the lowest median toxicity score (0.08). The low toxicity coupled with a low Gini score in group-II suggests consistent production of less toxic tweets. The wide range of Gini index values for groups III-VIII indicates the presence of both toxic and non-toxic tweets, with group-VIII consistently posting relatively more toxic tweets.

%-----%
\subsubsection{Lexical Diversity}
\label{sec:Lexical Diversity}
Effective communication improves understanding. We assess readability factors, including verbosity, grammar, semantics, and readability, across diverse thematic groups.
We examine tweet characteristics such as word count, sentence count, punctuation, and non-verbal cues (e.g., emoticons) to calculate Lexical Richness using the Automated Readability Index (ARI)~\cite{senter1967automated} and Flesch Score~\cite{flesch1948new}. Flesch Score gauges text complexity, while ARI evaluates comprehensibility. Higher Flesch and lexical diversity scores, along with lower ARI and Linsear scores, indicate better readability.
Table~\ref{tab:lexical_analysis} displays lexical metrics for each thematic group. Greater thematic diversity correlates with improved tweet comprehension. Group-VIII, with high diversity, exhibits optimal comprehension (highest Lexical diversity, lowest Linsear score). Conversely, lower diversity groups produce longer, less concise tweets.
Profiles with greater thematic diversity tend to use English more proficiently, resulting in shorter and more comprehensible tweets.
\begin{table}[!h]
\centering
%\scriptsize
\small
\tabcolsep=0.05cm
\resizebox{1\columnwidth}{!}{
\begin{tabular}{l|r|r|r|r|r|r|r}
\toprule
    & \multicolumn{1}{c}{\bf Gp-II}  & \multicolumn{1}{c}{\bf Gp-III}  & \multicolumn{1}{c}{\bf Gp-IV} & \multicolumn{1}{c}{\bf Gp-V} & \multicolumn{1}{c}{\bf Gp-VI}  & \multicolumn{1}{c}{\bf Gp-VII} & \multicolumn{1}{c}{\bf Gp-VIII}\\
            \hline
%  \midrule
\textbf{Flesch reading diff.}     &261.11&38.36&37.72&20.83&15.52&12.55&19.37\\
	\hline
	\textbf{Flesch reading ease}     &-612.85&-31.02&-26.34&13.63&29.62&44.49&37.06\\
	\hline
	\textbf{Linsear write score}     &472.37&61.87&59.73&28.1&18.95&15.26&28.34\\
	\hline
	\textbf{ARI score}     &343.81&50.73&51.09&29.35&21.40&16.60&24.56\\
	\hline
	\textbf{Lexical diversity} &19.97&29.83&42.83&47.01&92.01&119.1&128.1\\
	\hline
	\textbf{No. of chr per tweet} &108.0&102.7&103.3&107.8&110.8&85.4&61.0	\\
	\hline
	\textbf{No. of words per tweet} &18.0&14.3&14.7&14.0&15.0&12.9&10.6 \\
%\midrule
\bottomrule
\end{tabular}
}
\caption{\centering \small Overview of lexical diversity results for all II-VIII groups; all metric values are represented as average, calculated over all tweets of a profile (Section~\ref{sec:Lexical Diversity}).}
\vspace{-2mm}
\label{tab:lexical_analysis}
\end{table}
%-----%
\subsubsection{Prolificacy}
\label{sec:Prolificacy}
\begin{figure}[t!]
    \begin{subfigure}[b]{0.49\columnwidth}
    \centering
    \includegraphics[width=\columnwidth]{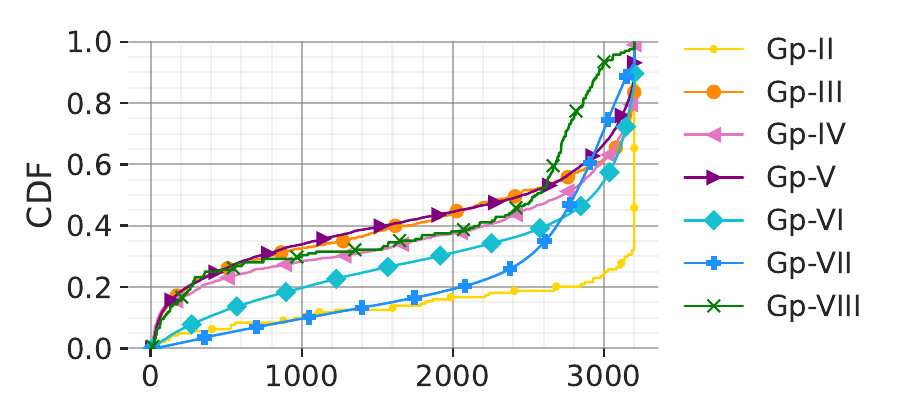}
    \caption{\# Tweets/profile}\label{fig:measurement_plots/no_tweets_per_profile.pdf}
    \end{subfigure}
    \begin{subfigure}[b]{0.49\columnwidth}
    \centering
    \includegraphics[width=\columnwidth]{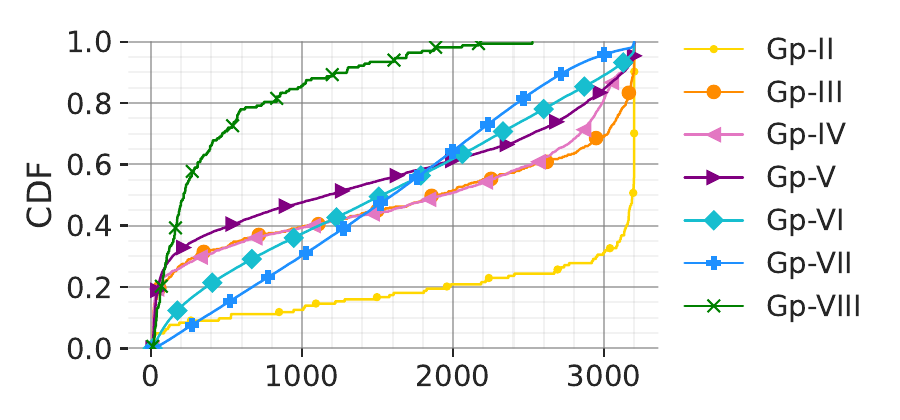}
    \caption{\# Unique tweets/profile}\label{fig:measurement_plots/no_unique_tweets_per_profile.pdf}
    \end{subfigure}
    \vspace{-2mm}
    \caption{\centering \small Cumulative distribution function of the number of total (\subref{fig:measurement_plots/no_tweets_per_profile.pdf}) and unique tweets (\subref{fig:measurement_plots/no_unique_tweets_per_profile.pdf}) per profile in groups II-VIII (Section~\ref{sec:Prolificacy}).}
    \label{fig:Number of tweets}
    \vspace{-5mm}
\end{figure}

Tweet count directly reflects Twitter profile activity. We calculate total and unique tweet numbers for all profile groups. Figure~\ref{fig:measurement_plots/no_tweets_per_profile.pdf} and Figure~\ref{fig:measurement_plots/no_unique_tweets_per_profile.pdf} present these figures for groups II-VIII. Profiles with higher thematic diversity show fewer tweets. Around 70\% of VII and VIII profiles have 3200 tweets, while groups VI and VII, and VIII show 2600-2800 tweets. More focused groups produce unique tweets, except for Group-III, where many have fewer unique tweets.
%
%-----%
\subsubsection{Hashtags}
\label{sec:Hashtags}
\begin{figure*}[!ht]
    \begin{subfigure}[b]{.33\textwidth}
    \centering
    \includegraphics[width=\textwidth]{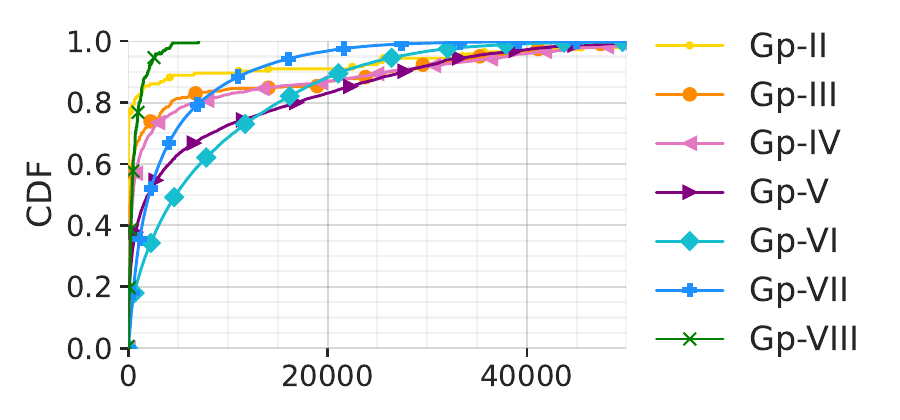}
    \caption{No. total hashtags}\label{fig:measurement_plots/total_hashtags_per_group.pdf}
    \end{subfigure}
    \begin{subfigure}[b]{.33\textwidth}
    \centering
    \includegraphics[width=\textwidth]{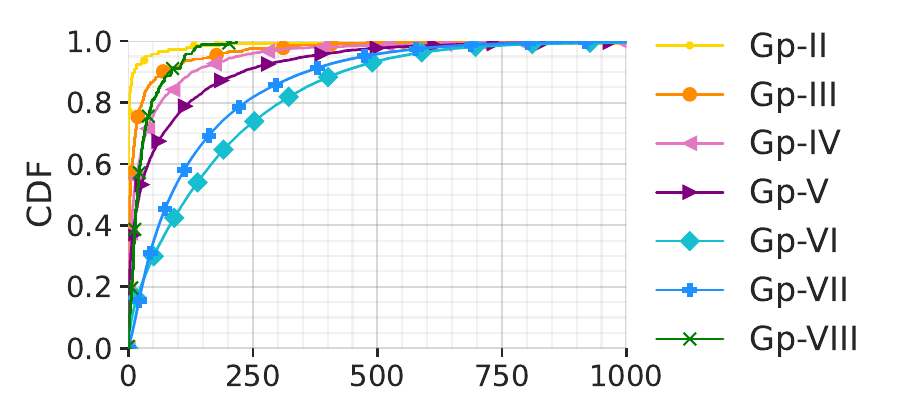}
    \caption{No. of unique hashtags}\label{fig:measurement_plots/unique_hashtags_per_group.pdf}
    \end{subfigure}
    \begin{subfigure}[b]{.33\textwidth}
    \centering
    \includegraphics[width=\textwidth]{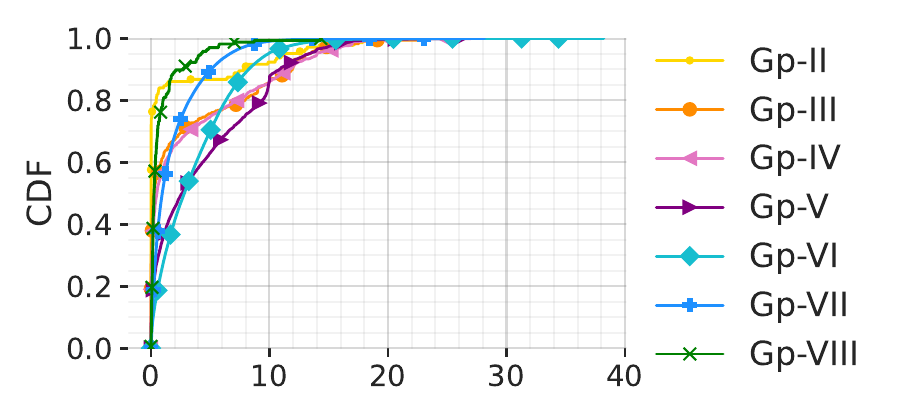}
    \caption{Ratio of hashtags to  tweets}\label{fig:measurement_plots/ratio_hashtags_to_ttw_all_groups.pdf}
    \end{subfigure}
    
    \vspace{-2mm}
    \caption{\centering \small (\subref{fig:measurement_plots/total_hashtags_per_group.pdf}) Cumulative distributions of the number of total hashtags; (\ref{fig:measurement_plots/unique_hashtags_per_group.pdf}) unique hashtags and; (\ref{fig:measurement_plots/ratio_hashtags_to_ttw_all_groups.pdf}) the ratio of hashtags to total the number of tweets  per profile in group-II to VIII (Section~\ref{sec:Hashtags}).}
    \label{fig:hashtags}
    \vspace{-4mm}
\end{figure*}

Hashtags offer insight into users' intended message communities. They've been used to detect profiles spreading rumors on Twitter~\cite{Addawood_Badawy_Lerman_Ferrara_2019, castillo2011information}. We analyze total hashtags, unique hashtags, and hashtag-to-tweet ratios for inauthentic profile indications.

Figure~\ref{fig:measurement_plots/total_hashtags_per_group.pdf} shows total hashtags per profile. Profiles in thematically diverse groups use more, with the top 30\% using 15K to 50K hashtags. Unique hashtags are in Figure~\ref{fig:measurement_plots/unique_hashtags_per_group.pdf}. More diverse profiles use fewer unique hashtags. The ratio of total hashtags to total tweets, in Figure~\ref{fig:measurement_plots/ratio_hashtags_to_ttw_all_groups.pdf}, indicates saturation. As groups increase, expected hashtags per tweet rise, except for group-VIII.

\subsection{Activity based analysis}
\label{sec:Activity based analysis}
\subsubsection{Burstiness}
\label{sec:Burstiness}
\begin{figure}[!ht]
    \centering
    \begin{subfigure}[b]{.70\columnwidth}
    \centering
    \includegraphics[width=\columnwidth]{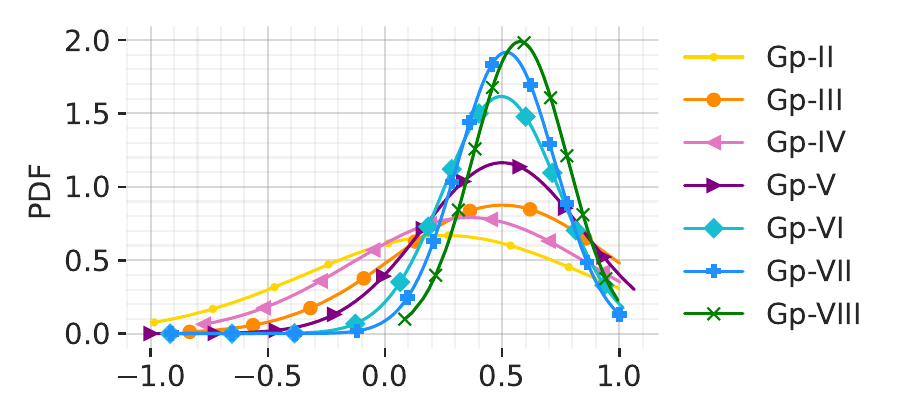}
    \caption{Normalized Burstiness}\label{fig:measurement_plots/normalised_burstiness_pdf.pdf}
    \end{subfigure}
     
    \begin{subfigure}[b]{.70\columnwidth}
    \centering
    \includegraphics[width=\columnwidth]{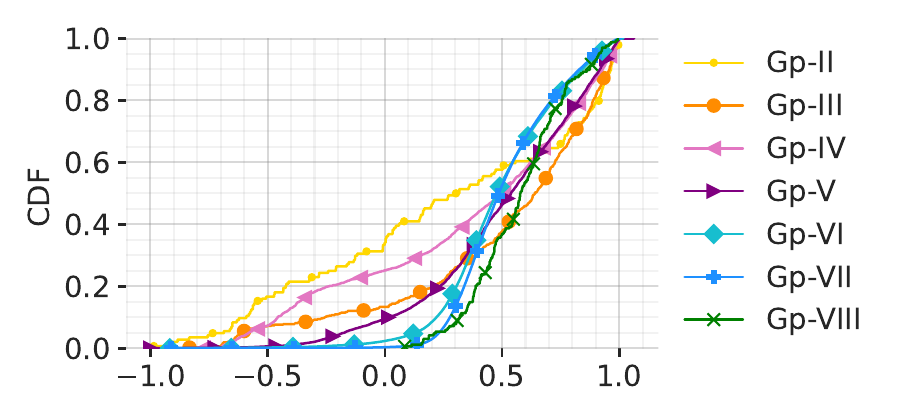}
    \caption{Normalized Burstiness}\label{fig:measurement_plots/normalised_burstiness_cdf.pdf}
    \end{subfigure}
    \vspace{-2mm}
    \caption{\centering \small Probability density function (\subref{fig:measurement_plots/normalised_burstiness_pdf.pdf}) and cumulative distribution function; (\subref{fig:measurement_plots/normalised_burstiness_cdf.pdf}) Plots of normalized burstiness score per profile in groups II-VIII (Section~\ref{sec:Burstiness}).}
    \label{fig:burstiness}
    % \vspace{-4mm}
\end{figure}

Exploring temporal tweeting patterns, we compare daily posting patterns of profiles in groups II-VIII through burstiness analysis~\cite{Kim_2016}. Burstiness characterizes inter-event times' distribution.
The \emph{Burstiness Score} ranges from -1 (periodic) to 1 (extremely bursty). Since we have a finite event count, we use \emph{Normalized Burstiness}~\cite{Kim_2016}, considering the topic time series of each group. 
Figure~\ref{fig:measurement_plots/normalised_burstiness_pdf.pdf} displays PDF and CDF of normalized burstiness. Higher thematically diverse profiles like VII, VIII tend to tweet randomly or without a pattern. In Figure~\ref{fig:measurement_plots/normalised_burstiness_cdf.pdf}, around 20\%-40\% of low diversity (e.g., groups II, IV) show highly regular tweeting, decreasing with more diversity (groups VI-VIII).
%
%-----%
\subsubsection{Daily tweeting pattern}
\label{sec:Daily tweeting pattern of profiles}
\begin{figure}[!ht]
    \centering
    \includegraphics[width=0.49\columnwidth]{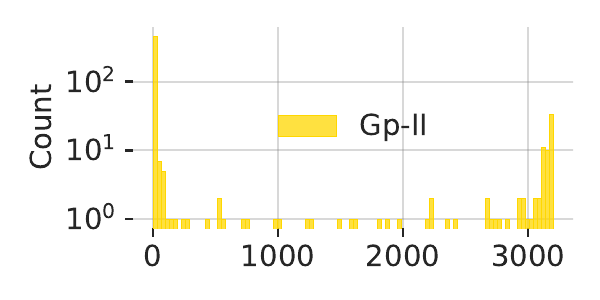}\includegraphics[width=0.49\columnwidth]{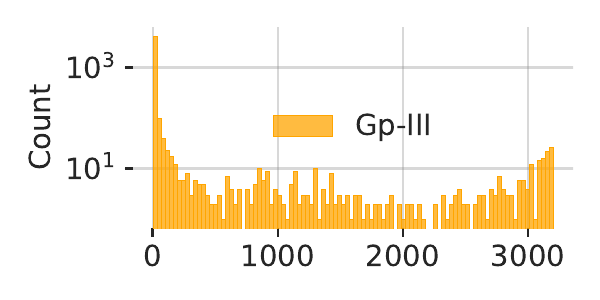}
    
    \includegraphics[width=0.49\columnwidth]{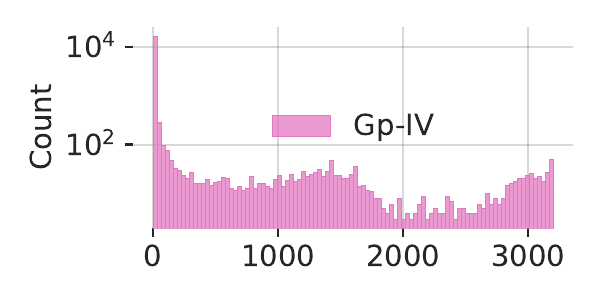}
    \includegraphics[width=0.49\columnwidth]{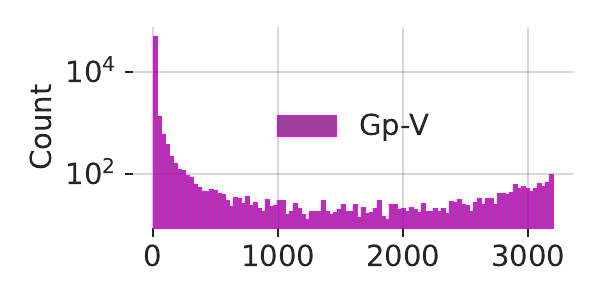}

    \includegraphics[width=0.49\columnwidth]{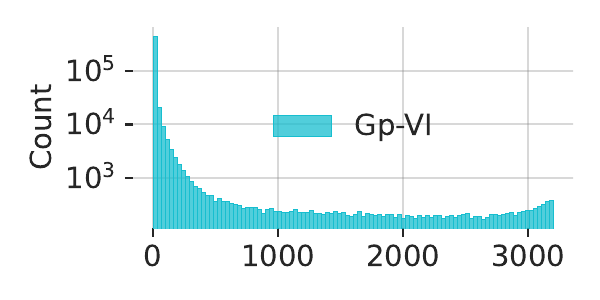}
    \includegraphics[width=0.49\columnwidth]{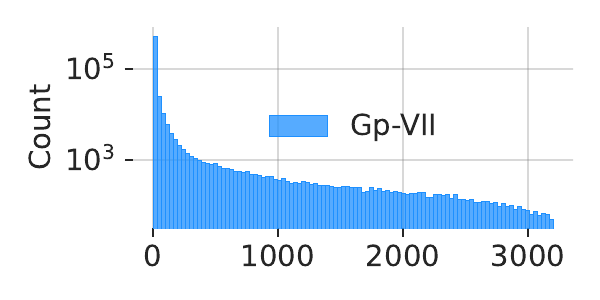}
    
    \includegraphics[width=0.49\columnwidth]{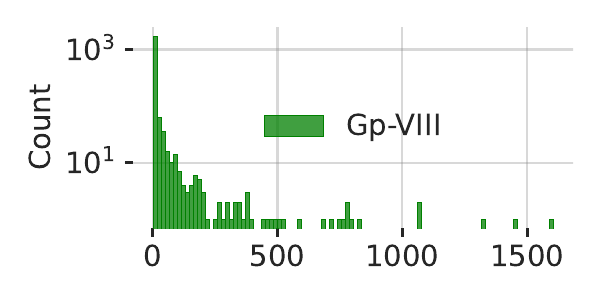}
    % \vspace{-2mm}
    \caption{\centering \small Tweeting patterns for profiles in groups II-VIII: Histograms displaying the day-to-day delta count between consecutive tweets (Section~\ref{sec:Daily tweeting pattern of profiles}).}
    \label{fig:tweeting_pattern}
    \vspace{-5mm}
\end{figure}
To further explore profile tweeting patterns, we delve into the original posting time of the tweets, represented by the tweet timestamps. Specifically, we examine the time intervals between consecutive tweets, known as the \emph{time delta}. Histograms illustrating these deltas (in days) for all profile groups can be found in Figure~\ref{fig:tweeting_pattern}. This figure portrays the distribution of the number of days between two successive tweets from each profile. Across all groups, the most common time delta is less than one day, signifying that a majority of Twitter profiles engage in daily tweeting activities. Notably, profiles in groups II, III, and VIII, representing both narrow-focused and highly diverse thematic profiles, tend to exhibit closely spaced tweet intervals of 1 to 100 days, whereas profiles in other groups display a more random posting pattern with varied time intervals. 
%-----%
\subsection{Profile based analysis}
\label{sec:Profile based analysis}
\subsubsection{Automation}
\label{sec:Automation}
\begin{table}[!]
\centering
\resizebox{0.99\columnwidth}{!}{
\begin{tabular}{l|c|c|c|c|c|c|c}
\toprule
{\bf Scr.}& \textbf{Gp-II} & \textbf{Gp-III} & \textbf{Gp-IV} & \textbf{Gp-V} & \textbf{Gp-VI} & \textbf{Gp-VII} & \textbf{Gp-VIII}\\ \midrule
\textbf{O}&  0.87$\pm$0.2& 0.71$\pm$0.3&   0.67$\pm$0.3&  0.60$\pm$0.3&  0.39$\pm$0.2&  0.28$\pm$0.2& 0.21$\pm$0.2\\ \hline
\textbf{S}&  0.43$\pm$0.1&      0.34$\pm$0.2&    0.33$\pm$0.2&  0.26$\pm$0.2&  0.08$\pm$0.1&  0.04$\pm$0.1& 0.08$\pm$0.1\\ 
\bottomrule
\end{tabular}
}
\caption{\centering \small Mean and Standard Deviation of Botometer scores including overall Bot score (O) and Spammer (S) score of profiles in group II-VIII (Section~\ref{sec:Automation}).}
\label{tab:new-botometer}
\vspace{-6mm}
\end{table}
The existence of automated accounts or bots on Twitter is well-established~\cite{TwitterBotAccounts}, and Twitter permits their use for legitimate purposes~\cite{Twitterautomation}. This study explores the relationship between the thematic diversity of content and profile automation.
To achieve this, we utilize the \emph{Botometer API v4}~\cite{Sayyadiharikandeh_2020}, which employs specialized classifiers to provide scores estimating the likelihood of different types of bots associated with a Twitter profile. These bot categories include self-declared bots, spammer bots, and more. Botometer assigns scores within the [0, 1] range, utilizing either English (incorporating all features) or Universal (language-independent) features. This study reports the universal feature scores. The interpretation of scores according to the API documentation is as follows:
\begin{itemize}
    \item {\bf Bot score}: A probabilistic score of the profile is a bot.
    \item {\bf Spammer}: A probabilistic score that a profile was labeled as spam bots from a number of datasets.
\end{itemize}
Table~\ref{tab:new-botometer} illustrates that bot-like behavior consistently diminishes as profiles exhibit greater thematic diversity in their tweets. The median bot score reduces from 0.87 in group II to 0.21 in group VIII. Furthermore, profiles in group II demonstrate the highest likelihood of being spamming, self-declared bots. Botometer results for our groups describe the proportion of bot-like and spamming profiles decreases with thematic diversity. Group-II has the lowest Botometer and Spammer scores.

%-----%
\subsubsection{Profile metadata}
\label{sec:Profile metadata}
Prior research on bot-like Twitter profiles~\cite{10.1145/3110025.3110090} has established that profile metadata, including factors like historic tweet count, retweets, and followers-to-following ratio, can effectively distinguish human profiles from bots. In addition to Botometer scores, we collected profile features across all eight thematic diversity groups.

To accomplish this, we used the Twitter API to retrieve the \emph{User object} containing account metadata like followers, followees, location, and profile name. We computed the followers-to-following ratio as an indicator of potential inauthentic profiles or bots, and the results are shown in Table~\ref{tab:profile_analysis}. Notably, the followers-to-following ratio consistently increases with higher thematic diversity, suggesting more human-like behavior in profiles.

Profiles involved in more discussion lists, especially those covering various themes, tend to be socially active and engage in Twitter communities, e.g., increased favorites and statuses.

\begin{figure}[t!] \centering
\includegraphics[width=0.95\columnwidth]{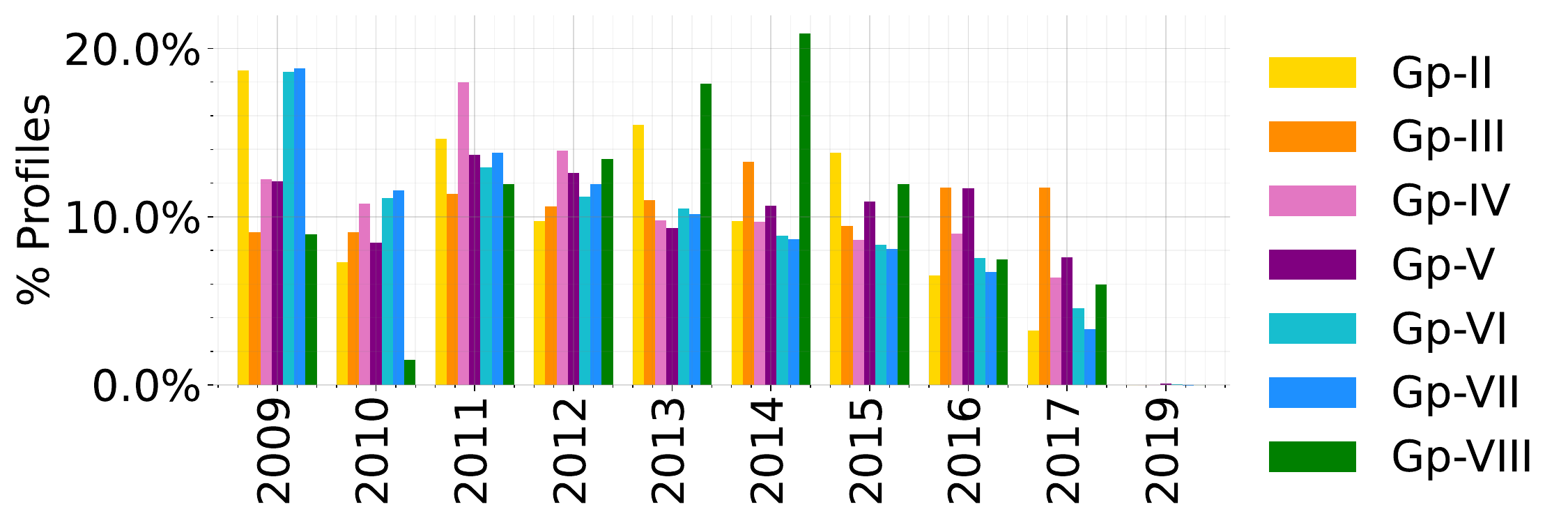}
\caption{\centering \small Barplot of the percentages of profiles created between 2007 and 2019 in groups II-VIII (Section~\ref{sec:Profile metadata}). }
\label{fig:profile_age_all_groups.pdf}
\vspace{-5mm}
\end{figure}

Regarding profile creation dates, Figure~\ref{fig:profile_age_all_groups.pdf} displays the distribution of profiles created from 2009 to 2019. Group VIII exhibits an imbalance in certain years like 2009, 2013, and 2014. Conversely, profiles with fewer thematic categories tend to be older, with a gradual decrease in newly created profiles from 2009 to 2017.
\begin{table*}[ht!]
\centering
%\scriptsize
\small
\tabcolsep=0.07cm
\resizebox{0.80\textwidth}{!}{
\begin{tabular}{l|c|c|c|c|c|c|c|c|r}
\toprule
\multicolumn{1}{l}{\bf Group}& 
   \multicolumn{1}{r}{\bf Followers}& 
   \multicolumn{1}{r}{\bf Following}&
   \multicolumn{1}{r}{\bf following /follower}&
   \multicolumn{1}{r}{\bf Listed}&
   \multicolumn{1}{r}{\bf Status}&
   \multicolumn{1}{r}{\bf Favourites}&
   \multicolumn{1}{r}{\bf Protected}&
   \multicolumn{1}{r}{\bf Verified}&
   \multicolumn{1}{r}{\bf Location}
   \\\hline
\textbf{\bf Gp-II}&40&10&0.25&3&3,375&2,002&1.62&1,020&UK(63.36\%)\\
	\hline
\textbf{\bf Gp-III}& 40& 11& 0.26& 3& 10,015& 0& 1.62& 3.63&UK(63.64\%) \\
    \hline
\textbf{\bf Gp-IV}&391&196&0.50&1&5,744&14,555&1.79&1.25&US(53.42\%) \\
    \hline
\textbf{\bf Gp-V}&642&384&0.59&5&15,786&2,060&2.53&10.84&US(47.10\%)\\
    \hline
\textbf{\bf Gp-VI }&3,232&842&0.26&44&13,021&6,021&2.08&23.32&US(49.1\%) \\
    \hline
\textbf{\bf Gp-VII}&1,536&790&0.51&17&18,346&12,720&4.58&14.57&US(52.8\%)\\
    \hline
\textbf{\bf Gp-VIII}& 626& 586& 0.93& 20&28,451& 22,923&8.95& 1.49& US(50.0\%)\\\hline
    % \hline
\bottomrule
\end{tabular}
}
\caption{\centering \small Profile metadata of the profiles in groups II-VIII. All values are represented as the average of all profiles in that group; only the follower/following column is presented as a ratio of avg following and follower number per group (Section~\ref{sec:Profile metadata}).}
\vspace{-2mm}
\label{tab:profile_analysis}
\end{table*}
%
%-----%
% \subsubsection{\bf Takeaway} 
% Thematically diverse Twitter accounts tweet unambiguously, their tweets are shorter and easily readable, although they do not have a specific posting pattern. They in general post toxic tweets when compared to profiles that tweet about a small number of themes or categories. Most thematically varied profiles in groups VI-VII are the most human-like profiles in our dataset with the lowest scores bot and spammer scores, with the group-VIII being the least. Also, these profiles have active and social Twitter accounts that like to interact with other Twitter profiles and be part of discussion communities. A high follower-to-following ratio supports the fact that they are popular profiles.
\section{Case Study: Thematically most diverse group (Group-VIII)} 
\label{sec:Case Study: Thematically most diverse group (Group-VIII)}
We establish that thematic diversity offers an intuitive means to differentiate between Twitter profiles in a meaningful manner, as evidenced by the substantial shifts observed in various content, activity, and profile-related trends. 
Building upon the insights garnered in Section~\ref{sec:Thematically diverse profile groups trends}, we identify group VIII as a collection of 168 profiles characterized by the highest level of thematic diversity and sustained toxicity. These profiles, ostensibly operated by humans, were predominantly created within a specific timeframe.
In pursuit of on-mission profiles, we adopt an approach that revisits the 200 individual topics assigned to each tweet, as detailed in Section~\ref{sec:Topic Modeling}. In this context, these topics provide a higher level of granularity compared to the previous eight thematic categories. While these categories previously served to gauge thematic diversity, they now enable us to achieve the specificity required for identifying cohesive narratives.

Furthermore, the topic modeling generates a Topic Probability Vector (TPV) for each tweet. By leveraging this vector, tweets can be attributed to one of the 200 topics. Consequently, the count of tweets associated with each individual topic facilitates the identification of the predominant topic or narrative contributed by each profile in the dataset. 
Moving forward, we leverage the predominant topic associated with each profile to uncover on-mission profiles within our dataset.

%-----%
\subsection{Topic label - predominating participation in a topic.}
\label{sec:Topic label}
The results of tweet-level topic modeling are represented as Topic Probability Vectors (TPVs) for each tweet within a profile. By averaging all the TPVs associated with a profile's tweets, we obtain an average TPV for that profile. To normalize this average TPV, we divide it by the global topic average score vector. The global topic average score vector is calculated by summing up the topic probabilities of all individual tweets (293 million) and then dividing by the total number of profiles (138,553). This normalization process produces a ``Normalized Topic Probability Vector" (nTPV):

\begin{equation}  
nTPV = \frac{\small \texttt{Average Topic Probability Score Vector}}{\small \texttt{Global Topic Average Score Vector}}
\label{eq:tpv}
\end{equation}

The nTPV helps identify profiles that exhibit above-average posting activity across all themes. Based on the highest topic probability in the nTPV, each profile is assigned a \emph{Topic label}. For instance, if a profile has the highest topic probability in topic \#10, it will be labeled as 10. The results of these topic labels can be found in Table~\ref{tab:on-mission profiles}, and for additional clarity, the topic categories obtained in Section~\ref{sec:Topic category} are presented.
\begin{table}[t!]
\centering
\resizebox{0.99\columnwidth}{!}{
{\large
\begin{tabular}{l|c|c|r}
\toprule
{\bf Gp-VIII}& {\bf Topic label}& {\bf Topic median tox.}& {\bf Topic cat.} \\
\midrule
62 (36.90\%)& 54&0.150	 & Politics \\\hline
26 (15.47\%)& 47& 0.148& Health \& covid \\\hline
8 (4.76\%)& 190& 0.146&News\& blogs \\ \hline
72 (42.85\%)& Misc.&0.09-0.08&Misc.\\
\bottomrule
\end{tabular}
}
}
\caption{\small \centering Profile groups associated with common topics, with the three largest communities presented. Approximately 42.56\% of profiles engaged in various topics, each group containing a maximum of 2 profiles (Section~\ref{sec:On-mission profiles})}
\vspace{-4mm}
\label{tab:on-mission profiles}
\end{table}
%-----%
\subsection{Topic toxicity}
\label{sec:Topic toxicity}
The topic label provides insight into the primary subject a profile tweets about the most. In this section, we delve into the details of our 200 topics, each comprised of tweets assigned to it based on the highest topic probability as determined by TPV (Section~\ref{sec:Topic Modeling}). To facilitate comparisons between topics, we compute the median toxicity value for each topic, using the toxicity scores of its constituent tweets.

To achieve this, we enhance each tweet with the \emph{Toxicity} score from \emph{Perspective API}~\cite{perspective}. This machine learning model assigns a probability score ranging from 0 to 1, indicating the level of toxicity present in a given tweet's content. Higher scores indicate higher toxicity. We employ the Perspective API to evaluate the toxicity score for all 293 million tweets in our dataset, allowing us to calculate the median toxicity for each topic. Our findings reveal that the median toxicity across all topics ranges from 0.08 to 0.15, highlighting variations in toxicity levels among different topics.
%-----%
\subsection{Topic labels of group-VIII profiles}
\label{sec:Topic label based profile grouping}
Resuming our examination of group-VIII profiles, we label each of the 168 profiles within this group with a specific topic label. This label uncovers the primary topic to which each profile has contributed the most out of the 200 topics. Furthermore, the median toxicity of the chosen topic provides insights into whether these profiles shared toxic tweets.
\subsubsection{Topic labeling}
\label{sec:Topic labeling}
The results of topic labeling for group-VIII profiles are presented in Table~\ref{tab:on-mission profiles}. Among these profiles, we identify 62, 26, and 8 profiles with the same topic labels \#54, \#47, and \#190, respectively. On the other hand, the remaining 72 profiles are typically associated with individual topics, often in isolation or small groups of two profiles at most. We delve deeper into the analysis of these profiles in the following sections.
%-----%
\subsubsection{Topic median toxicity}
The median toxicity of topics \#54, \#47, and \#190 are [0.150, 0.148, and 0.146] respectively. These values indicate a very high median toxicity, as evidenced by the upper quartile in Figure~\ref{fig:measurement_plots/TOXICITY_median_score_box_groups.pdf}. This suggests that tweets contributed by the 96 profiles to these topics have elevated toxicity scores. In contrast, the 72 profiles with individual topic labels displayed topic toxicity ranging from [0.09-0.08]. Based on this, we analyze the combined group of 96 profiles (62+26+8) and compare them with the remaining 72 profiles.
%-----%
\subsubsection{Network}
To accomplish this, we utilize the Twitter API to identify the friends of the aforementioned segregated profile groups. Our analysis reveals that 67\% of the 96-profile group are connected as friends, with 29\% of these pairs sharing over 15 friends on Twitter. Conversely, no such friend connections or shared friendships are observed within the 72-profile group.
%-----%
\subsubsection{Similar content}
Our investigation shows that within the 96-profile group, 29\% of the profiles engaged in retweeting similar content, accounting for 37\% of the retweets. Furthermore, among these connected profiles, there is a 32\% overlap in retweeted content with their respective individual followers. In contrast, no instances of shared retweets were observed among the 72-profile groups.
%-----%
\subsubsection{Variation in Category Participation Difference}

\begin{figure}[t!]
    \begin{subfigure}[b]{.49\columnwidth}
    \centering
    \includegraphics[width=\columnwidth]{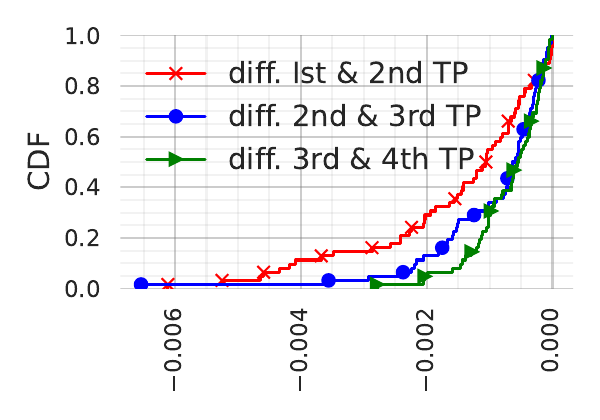}
    \caption{96 profiles}
    \label{fig:measurement_plots/diff_3_tps_politicalgroup.pdf}
    \end{subfigure}
    \begin{subfigure}[b]{.49\columnwidth}
    \centering
    \includegraphics[width=\columnwidth]{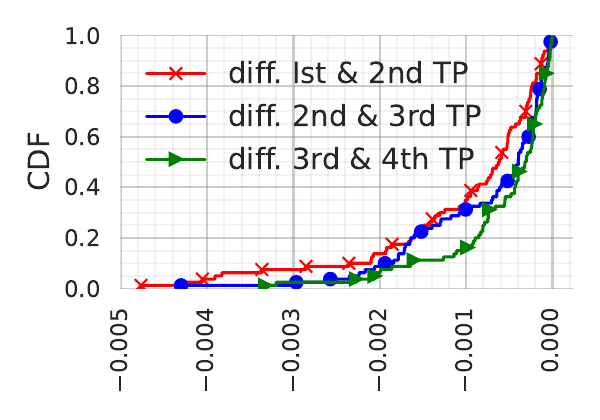}
    \caption{72 profiles}
    \label{fig:measurement_plots/diff_3_tps_baseline.pdf}
    \end{subfigure}
    \vspace{-1mm}
    \caption{\centering \small CDF of the difference in topic probabilities (volume of tweets) of top 3 topics of 96 (\subref{fig:measurement_plots/diff_3_tps_politicalgroup.pdf}) and 72 (\subref{fig:measurement_plots/diff_3_tps_baseline.pdf}) profiles }
    \label{fig:diff_tp_3topics}
    \vspace{-4mm}
\end{figure}

Given the considerable thematic diversity among all the profiles in group VIII, spanning across 8 distinct categories or themes, we aimed to delve into the discrepancies in their tweet volumes. Specifically, we examined the variation in tweet volumes between their dominant topics (identified through topic labels) and the subsequent two topics with the highest tweet counts.
As illustrated in Figure~\ref{fig:measurement_plots/diff_3_tps_politicalgroup.pdf} and Figure~\ref{fig:measurement_plots/diff_3_tps_baseline.pdf}, our analysis highlights that the disparity in probability scores between the first two topics is more pronounced within the 96-profile group when compared to the 72-profile group.
\subsubsection{On-mission profiles}
\label{sec:On-mission profiles}
In group-VIII, we identify two distinct profile clusters sharing similarities in thematic diversity, posting patterns, style, and human-like behavior. However, the 96-profile subset (51.75\%) primarily engages with a highly toxic topic, possessing content-sharing networks and followers who retweet their content. We classify this subset as ``on-mission" profiles, while the remaining 72 profiles (42.85\%) are labeled as ``not-on-mission" profiles. So:
\begin{quote}
    {\bf An ``on-mission" Twitter profile is characterized by its inclination to share a substantial number of toxic tweets focusing on a particular theme, which may encompass subjects such as ``politics," ``news," ``health," ``sports," and similar topics.}
\end{quote}

\section{Proposed detection model}
\label{sec:Proposed detection model}
In Section~\ref{sec:Topic label based profile grouping}, we identified 96 'on-mission' profiles. We now propose a detection model for finding similar profiles in groups II-VII. This model seeks to uncover shared attributes among on-mission profiles, regardless of specific topic features, enabling detection in new or emerging topics. The subsequent sections introduce features and classifiers.
%------%
\subsection{Features} 
\label{sec:Features}
Overall, we collect 38 different features, from on-mission and not-on-mission profiles. These features are categorized into three different categories:

\subsubsection{Content-based features} This category consists of features that elaborate the prolificacy of profiles in terms of 
(a){\it Tweet volume}:
(i) \#tweets in everyday, 
(ii) \#tweets in no topic, 
(iii) \#tweets in news/blogs, 
(iv) \#tweets in politics, 
(v) \#tweets in entertainment, 
(vi) \#tweets in sports, 
vii) \#tweets in profanity and 
(viii) \#tweets in health/covid. 
(b){\it Toxicity scores}: 
(i) median toxicity of all tweets per profile. 
(c) {\it Lexical diversity}:
(i) Flesch kincaid difficulty score, 
(ii) Flesch reading ease score, 
(iii) linear write score, 
(iv) automatic readability index,
(v) textual lexical diversity,
(d) {\it Tweet length}: 
(i) Number of characters per tweet, 
and (ii) the number of words per tweet.
\subsubsection{Auxiliary content based features} This set of features include:
(i) \#hashtags, 
(ii) \#unique hashtags, 
(iii) the number of hashtags to number of tweets ratio per profile, 
(iv) \#URLs, 
(v) \#unique URLs, 
and (vi) the ratio of URLs to tweets.
\subsubsection{Activity based features} 
This category of features shows the temporal behavior of profiles and consists of time deltas in days and the Normalized Burstiness of tweets. 
(a) The volume of content: 
(i) \#tweets,  
(ii) \#retweets, 
and (iii) \#unique tweets.
(b) Normalized Burstiness.
\subsubsection{\emph Profiles based features.} It also includes the meta-information of profiles and comprises 
(i)location,
(ii) description, 
(iii) protection level, 
(iv) followers count, 
(v) friends count, 
(vi) listed count, 
(vii) creation date, 
(viii) favorites count, 
(ix) geo-enabled, 
(x) verified, 
(xi) statuses count, 
(xii) contributors enabled, 
and (iii) withheld in countries.

\subsection{Proposed ML Models}

As elaborated in Section~\ref{sec:On-mission profiles}, we pinpointed 96 profiles under the on-mission category, alongside 72 profiles falling under the not-on-mission classification. To verify the precision of these designations, we carried out a manual assessment of both profile groups. 
% Below are instances of 5 consecutive tweets from an on-mission profile within topic \#54:
% \begin{enumerate}

% \item \texttt{\small ``Trump on Fox and Friends: “For three years I've been trying to figure out who is more corrupt- the New York Times or the Washington Post. When I find out I'll let you know.” Hosts do not object whatsoever.''}

% \item \texttt{\small``Will actually be buzzing if I can finish this assignment by Monday! Hallelujah!''}

% \item \texttt{\small ``Poignant and revealing deep-dive into the US commander in Afghanistan pardoned by Trump for 
% second degree murder? 
% TF I just read brah!''}

% \item \texttt{\small ``Trump Vocalized A Massive Truth About Terrorism ... But Is Anyone Paying Attention?''}

% \item \texttt{\small ``In secretly recorded audio, President Trump’s sister says he has ‘no principles’ and ‘you can’t trust him!!! TF'' }

% \item \texttt{\small ``My besties are booking their flights for the American final in Mexico. YuHUUU! wait-inggg!!!''}
% \end{enumerate}

The outcomes derived from our manual validation process form the foundation for training a machine learning classifier. This classifier is trained using features extracted from the identified on-mission and not-on-mission profiles. To build the supervised two-class classifiers, we leverage an open-source machine learning library called {\tt scikit-learn}~\cite{pedregosa2011scikit}. This library equips us with essential tools and algorithms for creating and deploying our classification models.
%-----%
\subsubsection{Validation} 
We utilize the {\tt MinMaxScaler()} function from the {\tt scikit-learn} library to normalize our three distinct feature categories for each profile group. Following normalization, the dataset is divided into three exclusive subsets: an 80\% training set and a 20\% test set. Four classifiers are trained using the supervised two-class support vector machines (SVM)~\cite{muller2018introduction} on the training set. The classifiers are structured to incorporate content, auxiliary, activity, and profile-based features individually, while the fourth classifier encompasses all features. Evaluation of the classifiers is performed using the ground-truth dataset of ``on-mission" and ``not-on-mission" profiles. Performance assessment involves analyzing F1-score ($\frac{2*TP}{2*TP+FP+FN}$) and accuracy ($\frac{TP+TN}{TP+TN+FP+FN}$) metrics\footnote{TP, TN, FP, and FN stand for true positives, true negatives, false positives, and false negatives, respectively.}. Our detection approach is also extended to other classification algorithms like Decision Tree~\cite{breiman2017classification} and Random Forests~\cite{breiman2001random}.

In Table~\ref{tab:classifier_validation}, the classifier results are presented, achieving F1-scores of 68.5\% to 100\% and accuracy rates of 67.7\% to 100\% in detecting on-mission profiles. Notably, Activity and Profile features exhibit high accuracy and F1-scores when paired with SVM and Random Forest classifiers. In contrast, Content-Based features demonstrate greater accuracy and F1-score alongside the Decision Tree classifier. The synergy of diverse feature categories in detecting on-mission profiles is evident, with the SVM classifier empirically displaying high accuracy and F1-score.
\begin{table}[t!]
\centering
\resizebox{1\columnwidth}{!}{

\tabcolsep=0.08cm
\begin{tabular}{l|rr|rr|rr}
\toprule
&\multicolumn{2}{c}{\bf SVM}  &\multicolumn{2}{c}{\bf Dec. Trees} &\multicolumn{2}{c}{\bf Rand. Forest}  \\
\cline{2-3}
\cline{3-5}
\cline{5-7}
%\cmidrule(r){2-4}
 
{\bf Features }& F1  & Acc. & F1  & Acc. & F1  & Acc. \\
\midrule
Content-Based	&	73.7\%	&	67.7\%	&	83.7\%	&	77.4\%	&	79.0\%	&	74.2\%	\\
Auxiliary	&	68.3\%	&	58.1\%	&	58.1\%	&	58.1\%	&	70.6\%	&	67.7\%	\\
Activity and Profile	&	86.5\%	&	83.9\%	&	64.7\%	&	61.3\%	&	83.3\%	&	80.7\%	\\\hline
All	&	100\%	&	100\%	&	100\%	&	100\%	&	94.1\%	&	93.6\%	\\
\bottomrule
\end{tabular}
}
\caption{\centering \small Classifier performance summary: F1-score and accuracy on our ground truth dataset}
\vspace{-6mm}
\label{tab:classifier_validation}
\end{table}
%
% \begin{table}[ht]
% \centering
% \resizebox{1\columnwidth}{!}{

% \tabcolsep=0.08cm
% \begin{tabular}{l|rr|rr|rr|rr}
% \toprule
% &\multicolumn{2}{c}{\bf SVM} &\multicolumn{2}{c}{\bf kNN}  &\multicolumn{2}{c}{\bf Dec. Trees} &\multicolumn{2}{c}{\bf Rand. Forest}  \\
% \cline{2-3}
% \cline{3-5}
% \cline{5-7}
% \cline{7-9}
% %\cmidrule(r){2-4}
 
% {\bf Features }& F1 (\%) & Acc. (\%) & F1 (\%) & Acc. (\%)& F1 (\%) & Acc. (\%)& F1 (\%) & Acc. (\%)\\
% \midrule
% Content Based	&	73.68	&	67.74	&	71.8	&	64.52	&	83.72	&	77.42	&	78.95	&	74.19	\\
% Auxiliary	&	68.29	&	58.06	&	78.95	&	74.19	&	58.06	&	58.06	&	70.59	&	67.74	\\
% Activity and Profile &	86.49	&	83.87	&	70.59	&	67.74	&	64.71	&	61.29	&	83.34	&	80.65	\\
% All	&	100	&	100	&	88.89	&	87.1	&	100	&	100	&	94.12	&	93.55	\\

% \bottomrule
% \end{tabular}
% }

% \caption{Summary of performance of classifiers. Here, we list the number of Twitter profiles participating in classifier training.}
% % \vspace{-5mm}
% \label{tab:classifier_validation}
% \end{table}
\subsection{Evaluation in the Wild}
We proceed to examine the two groups of profiles identified by our classifier as 'on-mission' and 'not-on-mission.' Our goal is to scrutinize the attributes of the 'on-mission' profiles flagged by the classifier and assess their similarity to on-mission or not-on-mission profiles in the ground truth. Utilizing our trained SVM model, we identify 'similar' profiles across all eight profile groups (Section~\ref{sec:Thematic diversity}). 
% within our dataset.

Table~\ref{tab:classifier_wild} presents the distribution and prevalence of flagged on-mission profiles across various groups, with percentages ranging from 44.39\% to 89.04\%, illustrating variations in the occurrence of flagged on-mission profiles among different profile groups.
%-----%
\subsubsection{Analysis of the classification results} 

In total, our classifier flagged 82,945 profiles as 'on-mission.' However, due to the labor-intensive nature of manual inspection, we adopted a pragmatic approach. For each entropy-based group (II-VIII), we randomly selected 100 profiles from both the flagged on-mission and not-on-mission groups, following the methodology in Section~\ref{sec:Case Study: Thematically most diverse group (Group-VIII)}. The outcomes of the process are in Table~\ref{tab:classifier_wild}.

From this table, among the 100 randomly sampled profiles from Groups III to VII, 89 were accurately identified as on-mission. However, all profiles from Groups III and VII were marked as not-on-mission, with only 26 profiles fitting this category. These statistics are based on a limited subset and don't represent the overall classifier performance.
We further examined the alignment between flagged on-mission profiles and ground-truth on-mission profiles, focusing on toxicity scores of their dominant topic and tweet volume discrepancies between predominant and other topics. We randomly chose 100 profiles from Groups II to VII, all flagged as on-mission by our classifier, and compared them to our ground-truth on-mission profiles (Section~\ref{sec:Thematically diverse profile groups trends}).

The findings in Table~\ref{tab:validation_of_manualchecking} show 373 (72\%) profiles annotated as on-mission and 164 (28\%) profiles as not-on-mission. Annotated on-mission profiles tend to engage with more toxic topics (based on median toxic topic scores) compared to others, as seen in Figure~\ref{fig:diff_tp_3topics}.
This observation validates that our annotated on-mission profiles are similar to ground-truth on-mission profiles in terms of topic toxicity and topic probability.
\begin{table}[t!]
\centering
\resizebox{1\columnwidth}{!}{
\tabcolsep=0.08cm
\begin{tabular}{l|r|r|c|c|c}
\toprule
&\multicolumn{5}{c}{\bf Profiles} \\
\cline{2-6}
%\cmidrule(r){2-4}
%  {\bf }& {\bf \# Available For} &  {\bf \# Flagged}\\
% {\bf Group }& {\bf Classification} & {\bf by ML Model} \\
% {\bf }&  &   & &\\
%&&&\multicolumn{2}{c}{\bf 10 Random Profiles} \\
%\cline{2-5}
{\bf Group }& {\bf \#Total } & {\bf \#Flagged} & {\bf \#Sample} & {\bf On Miss.} & {\bf Not-On-Miss. } \\
\midrule
Gp-II	&	32	&	20	(62.5\%)& {20}& {3}&{17}\\
Gp-III	&	223	&	100	(44.4\%)&  {100}&{21}&{79}\\
Gp-IV	&	1,289	&	614	(47.6\%)& {100}&{73}&{26}\\
Gp-V	&	3,175	&	1,812 (57.1\%)& {100}&{89}&{11}\\
Gp-VI	&	28,337	&	21,851 (77.1\%)& {100}&{83}&{17}\\
Gp-VII	&	65,758	&	58,549	(89.0\%)& {100}&{86}&{14}\\\hline
Total	&	98,814	&	82,945	(83.9\%)& {\bf 520}&%{\bf 373(72\%)}&{\bf 126(24\%)}\\
{\bf 373(72\%)}&{\bf 164(28\%)}\\

% On-Mission --> 

% On-not-mission --> 

%
\bottomrule
\end{tabular}

}
\caption{\centering \small SVM applied in the wild. Flagged Twitter profiles are manually annotated as on-mission and not-on-mission.}
\vspace{-5mm}
\label{tab:classifier_wild}
\end{table}
\begin{table}[h!]
\centering
\vspace{-4mm}
\resizebox{0.90\columnwidth}{!}{
{\large
\begin{tabular}{l|c|c|r}
\toprule
&\multicolumn{3}{c}{\bf Topic } \\
\cline{2-4}
{\bf Annot. On-miss.}& {\bf Label}& {\bf Median tox.}& {\bf Category} \\
\hline
155(41.5\%)& 10&0.140173 & Politics \\\hline
88(23.5\%)& 156& 0.142889& News/blogs \\\hline
69(18.4\%)& 32&0.126150 &  Politics \\\hline
17(\%)& 118& 	0.117828 &  News/blogs \\\hline
5(4.5\%)&100&0.136839 &  News/blogs \\\hline
39(10.4\%)& Unique&- & Multiple\\\hline
&\multicolumn{3}{c}{\bf Topic } \\
\cline{2-4}
{\bf Annot. Not-on-miss.}& {\bf Label}& {\bf Median tox.}& {\bf Category} \\
%{\bf }& {\bf }& {\bf }& {\bf } \\
\hline
57(45\%)& 124 & 0.092272 & News/blogs   \\\hline
17(\%)& 25&0.088606 &  Health/covid \\\hline
5(13.5\%)&166 &0.092408 &  Everyday \\\hline
3(2.38\%)& 1&0.104986 &  Health/covid \\\hline
44(34.9\%)& Unique&-&Multiple  \\

\bottomrule
\end{tabular}
}
}
\caption{\centering \small Topic scores and details of the topics of the 520 profiles manually annotated as on-mission and not-on-mission. }
\vspace{3mm}
\label{tab:validation_of_manualchecking}
\vspace{-5mm}
\end{table}
\begin{figure}[h!]
    \begin{subfigure}[b]{.49\columnwidth}
    \centering
    \includegraphics[width=\columnwidth]{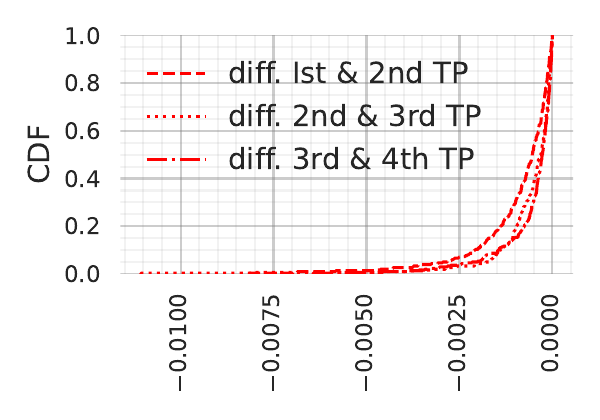}
    \caption{Annotated On-mission.}\label{fig:measurement_plots/diff_3_tps_on_mission_annot.pdf}
    \end{subfigure}
    \begin{subfigure}[b]{.49
\columnwidth}
    \centering
    \includegraphics[width=\columnwidth]{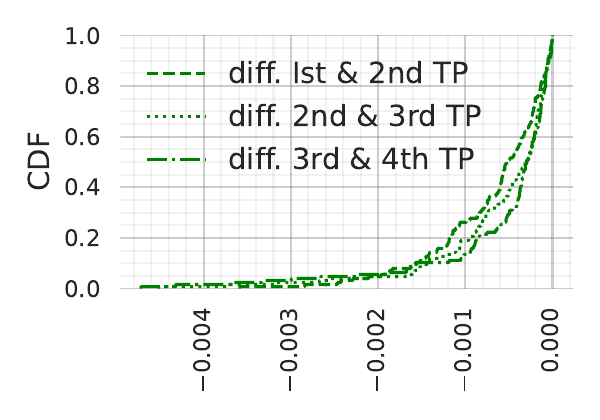}
    \caption{Annotated Not on-mission. }\label{fig:measurement_plots/diff_3_tps_not_on_mission_annot.pdf}
    \end{subfigure}
    \vspace{2mm}
    \caption{\centering \small Validation of detection method in the wild. We present the difference between the top 3 topic probabilities of manual validation of flagged profiles: on-mission vs. not-on-mission. }
    \label{fig:diff_tp_3topics_manual_annot}
    \vspace{-4mm}
\end{figure}
\section{Ethical Considerations}
\label{sec:Ethical Considerations}
Macquarie University IRB Project Reference: 35379, Project ID: 10008, Granted: 27/11/2021.

% This study is non-commercial study, we strictly adhere to ethical guidelines as outlined in~\cite{rivers2014ethical}, and obtained an ethics committee approval from our institution's IRB due to experimentation with human-generated data.
% Our data will not be used for any commercial purposes and will remain private. 
% \footnote{Macquarie University IRB Project Reference: 35379, Project ID: 10008, Granted: 27/11/2021}

\section{Conclusion}
\label{sec:Discussion and conclusion}
In today's globalized era, social media offers cost-effective tools for malicious actors, reducing the need for physical involvement or large advertising budgets in polarizing society. These platforms facilitate influence operations, often driven by manipulated profiles, whether automated or human-operated. While the full impact of this manipulation on preferences is not fully understood, it's a recognized concern among academics and authorities.

Our findings reveal that on-mission profiles strategically use thematic content diversity to build trust among users. Additionally, distinctions between on-mission profiles and genuine profiles on Twitter can be discerned through content.

Detectable thematic diversity on Twitter provides insights into its dynamics. However, profiles engaging in targeted toxic behavior directed at specific themes are likely 'on-mission,' focusing on propagating polarizing and toxic tweets related to topics like politics, health, sports, news, and COVID-19.

These profiles frequently share easily understandable content rapidly, with intermittent periods of dormancy. They also accumulate a substantial number of followers, suggesting human management rather than automated systems.

Distinguishing participant profiles from regular ones is crucial in addressing real-time influence operations. Tools like Botometer~\cite{Sayyadiharikandeh_2020} can help identify less sophisticated influence attempts or promotional operations relying on a limited number of bot-operated accounts. Detecting profiles used in complex influence operations involving a combination of human and hybrid-operated accounts is more challenging. Developing detection systems independent of platform-specific features is vital, as foreign agents exploit various social media platforms.
Our approach is expected to perform well in recognizing on-mission profiles in new campaigns because classifiers trained on data from known campaigns struggle to identify tactics and features.

% Future research will focus on cross-platform tracking of influence operations over extended periods, identifying and analyzing coordinated campaigns across diverse social media platforms.

% \bibliography{plain}
% \bibliographystyle{ACM-Reference-Format}

% \setlength{\bibitemsep}{0.5em} % Adjust the spacing as needed
\bibliographystyle{plain}
% \bibliography{references}

% \appendix
% \input{appendix}

\end{document}